\documentclass[conference]{IEEEtran}
\IEEEoverridecommandlockouts
\usepackage{cite}
\usepackage{amsmath,amssymb,amsfonts}
\usepackage{algorithmic}
\usepackage{graphicx}
\usepackage{textcomp}
\usepackage{xcolor}
\def\BibTeX{{\rm B\kern-.05em{\sc i\kern-.025em b}\kern-.08em
    T\kern-.1667em\lower.7ex\hbox{E}\kern-.125emX}}

\usepackage{cite}
\usepackage{amsmath,amssymb,amsfonts,bbm}
\usepackage{graphicx}
\usepackage{textcomp}
\usepackage[dvipsnames]{xcolor}
\usepackage{multirow}
\usepackage{subcaption}
\usepackage{caption}\usepackage{tikz}
\usepackage{tkz-tab}
\usetikzlibrary{automata,arrows,positioning,calc}
\usetikzlibrary{shapes,snakes}
\usepackage{dsfont}
\usepackage{soul}
\usepackage{subfiles}
\usepackage{comment}
\usepackage{amsthm}
\usepackage{booktabs}
\usepackage{epstopdf} %converting to PDF
\usepackage{hyperref}
\hypersetup{
    hidelinks
}

\usepackage{tikz}
\usetikzlibrary{arrows.meta,shapes.geometric,positioning,fit,backgrounds}
\usepackage{pgfplots}
\pgfplotsset{compat=1.18}
\usepackage{url}
\usepackage{listings}
\usepackage{microtype}

\usepackage{pifont}% http://ctan.org/pkg/pifont
\newcommand{\cmark}{\ding{51}}%
\newcommand{\xmark}{\ding{55}}%

\usepackage[acronym]{glossaries}
\makeglossaries
\newacronym{ieee}{IEEE}{Institute of Electrical and Electronics Engineers}
\newacronym{vr}{VR}{Virtual Reality}
\newacronym{ar}{AR}{Augmented Reality}
\newacronym{obss}{OBSS}{Overlapping Basic Service Set}
\newacronym{iot}{IoT}{Internet of Things}
\newacronym[plural=APs]{ap}{AP}{Access Point}
\newacronym{csmaca}{CSMA/CA}{Carrier Sense Multiple Access with Collision Avoidance}
\newacronym{lbt}{LBT}{Listen-Before-Talk}
\newacronym[plural=BSSs]{bss}{BSS}{Basic Service Set}
\newacronym{mlo}{MLO}{Multi-Link Operation}
\newacronym{sr}{SR}{Spatial Reuse}
\newacronym{obsspd}{OBSS/PD}{Overlapping Basic Service Set/Packet Detect}
\newacronym[plural=STAs]{sta}{STA}{Station}
\newacronym[plural=TXOPs]{txop}{TXOP}{Transmit Opportunities}
\newacronym{snr}{SNR}{Signal-to-Noise-Ratio}
\newacronym{tgbn}{TGbn}{Task Group bn}
\newacronym{cosr}{Co-SR}{Coordinated Spatial Reuse}
\newacronym{cobf}{Co-BF}{Coordinated Beamforming}
\newacronym{cortwt}{Co-RTWT}{Coordinated R-TWT}
\newacronym{cotdma}{Co-TDMA}{Coordinated Time Division Multiple Access}
\newacronym{jt}{JT}{Joint Transmissions}
\newacronym{coofdma}{Co-OFDMA}{Coordinated OFDMA}
\newacronym{mapc}{MAPC}{Multi-Access Point Coordination}
\newacronym{ecsr}{ECSR}{Enhanced Coordinated Spatial Reuse}
\newacronym{rssi}{RSSI}{Received Signal Strength Indicator}
\newacronym{mcs}{MCS}{Modulation and Coding Scheme}
\newacronym{dcf}{DCF}{Distributed Coordination Function}
\newacronym{sinr}{SINR}{Signal-to-Interference-plus-Noise Ratio}
\newacronym{rrm}{RRM}{Radio Resource Management}
\newacronym{rts}{RTS}{Request-to-Send}
\newacronym{cts}{CTS}{Clear-to-Send}
\newacronym{arq}{ARQ}{Automatic Repeat Request}
\newacronym{ampdu}{A-MPDU}{Aggregate MAC Protocol Data Unit}
\newacronym{ai}{AI}{Artificial Intelligence}
\newacronym{ml}{ML}{Machine Learning}
\newacronym{phy}{PHY}{Physical}
\newacronym{mac}{MAC}{Medium Access Control}
\newacronym{uhr}{UHR}{Ultra-High Reliability}
\newacronym{sifs}{SIFS}{Short Interframe Space}
\newacronym{difs}{DIFS}{DCF Interframe Space}
\newacronym{aifs}{AIFS}{Arbitration Interframe Space}
\newacronym{wds}{WDS}{Wireless Distribution System}
\newacronym{ssid}{SSID}{Service Set Identifier}
\newacronym{ess}{ESS}{Extended Service Set}
\newacronym{mbss}{MBSS}{Mesh Basic Service Set}
\newacronym{wlc}{WLC}{Wireless LAN Controller}
\newacronym{wlan}{WLAN}{Wireless Local Area Network}
\newacronym{capwap}{CAPWAP}{Control And Provisioning of Wireless Access Points}
\newacronym{ietf}{IETF}{Internet Engineering Task Force}
\newacronym{rfc}{RFC}{Request for Comments}
\newacronym{isp}{ISP}{Internet Service Provider}
\newacronym{cpe}{CPE}{Customer-Premises Equipment}
\newacronym{usp}{USP}{User Services Platform}
\newacronym{mqtt}{MQTT}{Message Queuing Telemetry Transport}
\newacronym{bssid}{BSSID}{Basic Service Set Identifier}
\newacronym{tpc}{TPC}{Transmit Power Control}
\newacronym{cocr}{Co-CR}{Coordinated Channel Recommendation}
\newacronym{pasn}{PASN}{Pre-Association Security Negotiation}
\newacronym{ltf}{LTF}{Long Training Field}
\newacronym{gi}{GI}{Guard Interval}
\newacronym{twt}{TWT}{Target Wake Time}
\newacronym{sp}{SP}{Service Period}
\newacronym{rtwt}{R-TWT}{Restricted Target Wake Time}
\newacronym{aid}{AID}{Association ID}
\newacronym{pmksa}{PMKSA}{Pairwise Master Key Security Association}
\newacronym{ptksa}{PTKSA}{Pairwise Transient Key Security Association}
\newacronym{rsne}{RSNE}{Robust Security Network Element}
\newacronym{rsnxe}{RSNXE}{RSN eXtension Element}
\newacronym{mic}{MIC}{Message Integrity Check}
\newacronym{back}{BACK}{Block Acknowledgment}
\newacronym{ppdu}{PPDU}{Physical Layer Protocol Data Unit}
\newacronym{csi}{CSI}{Channel State Information}
\newacronym{ndp}{NDP}{Null Data Packet}
\newacronym{bsrp}{BSRP}{Buffer Status Report Poll}
\newacronym{ntb}{NTB}{Non-Trigger Based}
\newacronym{icf}{ICF}{Initial Control Frame}
\newacronym{icr}{ICR}{Initial Control Response}
\newacronym{dps}{DPS}{Dynamic Power Save}
\newacronym{emlsr}{EMLSR}{Enhanced Multi-Link Single Radio}
\newacronym{ofdm}{OFDM}{Orthogonal Frequency-Division Multiplexing}
\newacronym{eht}{EHT}{Extremely High Throughput}
\newacronym{ndpa}{NDPA}{NDP Announcement}
\newacronym{tb}{TB}{Trigger-Based}
\newacronym{bfrp}{BFRP}{Beamforming Report Poll}
\newacronym{ofdma}{OFDMA}{Orthogonal Frequency-Division Multiple Access}
\newacronym{mu}{MU}{Multi-User}
\newacronym{txs}{TXS}{Triggered TXOP Sharing}
\newacronym{mpdu}{MPDU}{MAC Protocol Data Unit}
\newacronym{tsf}{TSF}{Time Synchronization Function}
\newacronym{p2p}{P2P}{Peer-to-Peer}
\newacronym{txspg}{TXSPG}{TXOP Sharing for P2P Groups}
\newacronym{scs}{SCS}{Stream Classification Service}
\newacronym{qos}{QoS}{Quality of Service}
\newacronym{ru}{RU}{Resource Unit}
\newacronym{mimo}{MIMO}{Multiple Input Multiple Output}
\newacronym{cst}{CST}{Carrier Sense Threshold}
\newacronym{irm}{IRM}{Internal Regret Minimization}
\newacronym{ce}{CE}{Correlated Equilibrium}
\newacronym{ne}{NE}{Nash Equilibrium}
\newacronym{cca}{CCA}{Clear Channel Assessment}
\newacronym{psr}{PSR}{Parametrized Spatial Reuse}
\newacronym{mab}{MAB}{Multi-Armed Bandit}
\newacronym{ma}{MA}{Multi-Agent}
\newacronym{drl}{DRL}{Deep Reinforcement Learning}
\newacronym{rl}{RL}{Reinforcement Learning}
\newacronym{npca}{NPCA}{Non-Primary Channel Access}
\newacronym{dso}{DSO}{Dynamic Subband Operation}
\newacronym{edca}{EDCA}{Enhanced Distributed Channel Access}
\newacronym{dru}{DRU}{Distributed Resource Units}
\newacronym{fsm}{FSM}{Finite State Machine}
\newacronym{ack}{ACK}{acknowledgment}
\newacronym{nav}{NAV}{Network Allocation Vector}
\newacronym{tf}{TF}{Trigger Frame}
\newacronym{ula}{ULA}{Uniform Linear Array}
\newacronym{zf}{ZF}{Zero Forcing}
\newacronym{beb}{BEB}{Binary Exponential Backoff}
\newacronym{vo}{VO}{Voice}
\newacronym{vi}{VI}{Video}
\newacronym{be}{BE}{Best Effort}
\newacronym{bk}{BK}{Background}
\newacronym[plural=ACs]{ac}{AC}{Access Categorie}
\newacronym{cw}{CW}{Contention Window}
\newacronym{iyt}{IYT}{It's Your Turn}
\newacronym{eca}{ECA}{Enhanced Collision Avoidance}
\newacronym{dcb}{DCB}{Dynamic Channel Bonding}
\newacronym{dnn}{DNN}{Deep Neural Network}
\newacronym{gnn}{GNN}{Graph Neural Network}
\newacronym{psm}{PSM}{Power Save Mode}
\newacronym{dqn}{DQN}{Deep Q-Network}
\newacronym{ppo}{PPO}{Proximal Policy Optimization}
\newacronym{maddpg}{MADDPG}{Multi-Agent Deep Deterministic Policy Gradient}
\newacronym{murts}{MU-RTS}{Multi-User Request-to-Send}

\usepackage[most]{tcolorbox}

\newtcblisting{bashbox}[2][]{
	colback=gray!5!white,          
	colframe=gray!75!black,        
	listing engine=listings,       % <-- Changed to listings
	listing only,                  
	listing options={              % <-- Replaced minted options
		language=bash,
		breaklines=true, 
		basicstyle=\footnotesize\ttfamily,
		showstringspaces=false
	},
	title={#2},                    
	fonttitle=\small\bfseries\ttfamily,
	enhanced,
	attach boxed title to top left={yshift=-4.5mm, xshift=2mm},
	bottom=-3mm,      
	boxed title style={colback=gray!75!black},
	#1                             
}

\begin{document}

% This code is to reduce the list of authors by using et. al:
\bstctlcite{IEEEexample:BSTcontrol}

\title{\texttt{Kom8ndor}: An IEEE 802.11bn-Oriented Simulator for Wi-Fi 8 and Beyond
\thanks{This work was supported by the following projects: TRUE Wi-Fi PID2024-155470NB-I00 (MICIU/AEI/10,13039/501100011033/FEDER,UE), MLDR (Chist-ERA WAI 2022) PCI2023-145958-2 (MCIU/AEI/10.13039), ICREA Academia 2024 (00077 AGAUR), and MdM CEX2021-001195-M (MICIU/AEI/10.13039/501100011033).}
}

\author{\IEEEauthorblockN{Francesc Wilhelmi}
\IEEEauthorblockA{\textit{Engineering Department} \\
\textit{Universitat Pompeu Fabra}\\
Barcelona, Spain \\
francisco.wilhelmi@upf.edu}
\and
\IEEEauthorblockN{Sergio Barrachina-Muñoz}
\IEEEauthorblockA{\textit{Services As NetworkS} \\
\textit{Centre Tecnològic de Telecomunicacions de Catalunya}\\
Barcelona, Spain \\
sergio.barrachina@cttc.cat}
\and
\IEEEauthorblockN{Boris Bellalta}
\IEEEauthorblockA{\textit{Engineering Department} \\
\textit{Universitat Pompeu Fabra}\\
Barcelona, Spain \\
boris.bellalta@upf.edu}
}

\maketitle

%%%%%%%%%%
% FORCE PAGE NUMBERS
\thispagestyle{plain}
\pagestyle{plain}
%%%%%%%

\begin{abstract}
The upcoming IEEE 802.11bn amendment marks a paradigm shift in Wi-Fi, which will pose ambitious performance targets under the paradigm of Ultra-High Reliability (UHR). To understand the implications of such a new technology and to support early research and protocol design for Wi-Fi~8, we present \texttt{Kom8ndor}. This discrete-event network simulator extends the open-source Komondor platform (a simulator validated against ns-3 and other analytical tools) with 802.11bn features. Among the newly added functionalities, we highlight Multi-Access Point Coordination (MAPC)---including Coordinated Time-Division Multiple Access (Co-TDMA), Coordinated Spatial Reuse (Co-SR), and Coordinated Beamforming (Co-BF)---, Non-Primary Channel Access (NPCA), and Dynamic Subband Operation (DSO). Beyond Wi-Fi~8 implementations, \texttt{Kom8ndor} introduces novel functionalities (e.g., a machine learning wrapper for building AI-based protocols) and a modular design to boost the prototyping and research of future Wi-Fi technologies. \texttt{Kom8ndor} is open-source (GNU GPLv3) and available at \url{https://github.com/wn-upf/Komondor}.
\end{abstract}


\section{Introduction}
\label{sec:introduction}

With the upcoming IEEE 802.11bn-2028 (Wi-Fi 8), \glspl{wlan} are set to go beyond supporting mere Internet access, and step toward communications-critical applications like industrial automation~\cite{geraci2026wi}. To meet \gls{uhr} goals, 802.11bn will introduce new functionalities, such as \gls{npca}, \gls{dso}, or \gls{mapc}. Given the unprecedented enhancements and architectural changes that 802.11bn proposes over 802.11, the need for open-source simulation tools becomes critical, as they enable a thorough evaluation of both standard functionalities and novel extensions (e.g., beyond 802.11bn \gls{mapc}) in various scenarios and use cases. 

ns-3~\cite{riley2010ns} remains the most popular choice, offering high-fidelity, full-stack simulation with a mature set of Wi-Fi 7 features (e.g., \gls{mlo}, \gls{mu}-\gls{ru} \gls{ofdma}), while OMNeT++/INET~\cite{varga2008overview} provides high modularity and realistic emulation, and MATLAB's WLAN Toolbox~\cite{matlab_wlan_toolbox} excels at high-fidelity link-level and \gls{phy} modeling (including some Wi-Fi 8 \gls{phy} models). All three, however, follow a slow but necessary development pace that limits rapid prototyping of fast-evolving standards, pushing early Wi-Fi research efforts toward narrow-scope (sometimes closed-source) tools~\cite{carrascosa2024performance, alsakati2023performance, jeknic2023development, chen2022overview}. Regarding 802.11bn, very few works provide simulation results given the technology's early status (currently Draft 2.0), and existing examples are either not standard-compliant and tightly coupled to specific proposals~\cite{nunez2025enabling} or not open-sourced~\cite{liu2024wi}. To the best of our knowledge, no publicly available simulator offers Wi-Fi~8 and beyond functionalities within an extensible framework (see Table~\ref{tab:simulators}).

\begin{table}[t!]
\caption{High-level comparison of Wi-Fi simulation tools.}
\label{tab:simulators}
\resizebox{\columnwidth}{!}{%
\begin{tabular}{@{}lccccc@{}}
\toprule
\multicolumn{1}{c}{\textbf{Tool}} & \textbf{\begin{tabular}[c]{@{}c@{}}Open-\\ source\end{tabular}} & \textbf{Main strength} & \textbf{Wi-Fi 8} & \multicolumn{1}{c}{\textbf{\begin{tabular}[c]{@{}c@{}}Future \\ Wi-Fi\end{tabular}}} \\ \midrule
ns-3 & \cmark & \begin{tabular}[c]{@{}c@{}}Full-stack, \\ wide adoption.\end{tabular} & \xmark & \xmark \\ \midrule
\begin{tabular}[c]{@{}l@{}}MATLAB's \\ WLAN Toolbox\end{tabular} & \xmark & \begin{tabular}[c]{@{}c@{}}High-fidelity \\ PHY modeling.\end{tabular} & \cmark$^a$ & \xmark \\ \midrule
\begin{tabular}[c]{@{}l@{}}OMNeT++/\\ INET\end{tabular} & \cmark & \begin{tabular}[c]{@{}c@{}}High modularity, \\ realistic emulation.\end{tabular} & \xmark & \xmark \\ \midrule
\texttt{Kom8ndor} & \cmark & \begin{tabular}[c]{@{}c@{}}Fast event processing, \\ rapid prototyping, \\ AI/ML integration.\end{tabular} & \cmark$^b$ & \cmark \\ \bottomrule
\end{tabular}%
}
\footnotesize{$^a$Includes Wi-Fi 8 PHY models (e.g., new MCSs), $^b$Built explicitly to evaluate MAPC and other Wi-Fi 8 features.}\\
\end{table}

For this reason, we introduce \texttt{Kom8ndor}, a C/C++ 802.11bn-oriented event-driven network simulator that substantially extends the original Komondor simulator \cite{barrachina2019komondor} (originally meant for Wi-Fi~6 simulations~\cite{wilhelmi2021spatial, barrachina2019dynamic}) to introduce Wi-Fi~8 and beyond features, including a new \gls{mapc} framework alongside \gls{dso}, \gls{npca}, and new channel access methods. \texttt{Komondor} serves as a well-suited simulation platform for Wi-Fi research, offering a balance between implementation speed and simulation fidelity that more comprehensive frameworks like ns-3 are less suited to provide for rapidly evolving standards, while retaining a lightweight, high-throughput \gls{mac}/\gls{phy} abstraction well suited to rapid prototyping. Moreover, \texttt{Kom8ndor} extends the \gls{ml} modules from Komondor by introducing a Python wrapper, which is key to accelerating the experimentation of AI-native solutions for future Wi-Fi.

\section{Architecture and Design}
\label{sec:architecture_design}

Komondor was initially designed for cost-effective simulation of high-density Wi-Fi 6 scenarios, featuring novel features at that time, such as \gls{obsspd}-based \gls{sr}~\cite{wilhelmi2021spatial} or Dynamic Channel Bonding (DCB)~\cite{barrachina2019dynamic}. \texttt{Kom8ndor} extends the original architecture to introduce Wi-Fi 8 and beyond features, higher modularity for forward extensions and integration of third-party \gls{ml} tools, higher efficiency, and better usability.

\begin{figure}[t!]
  \centering
  \includegraphics[width=\columnwidth]{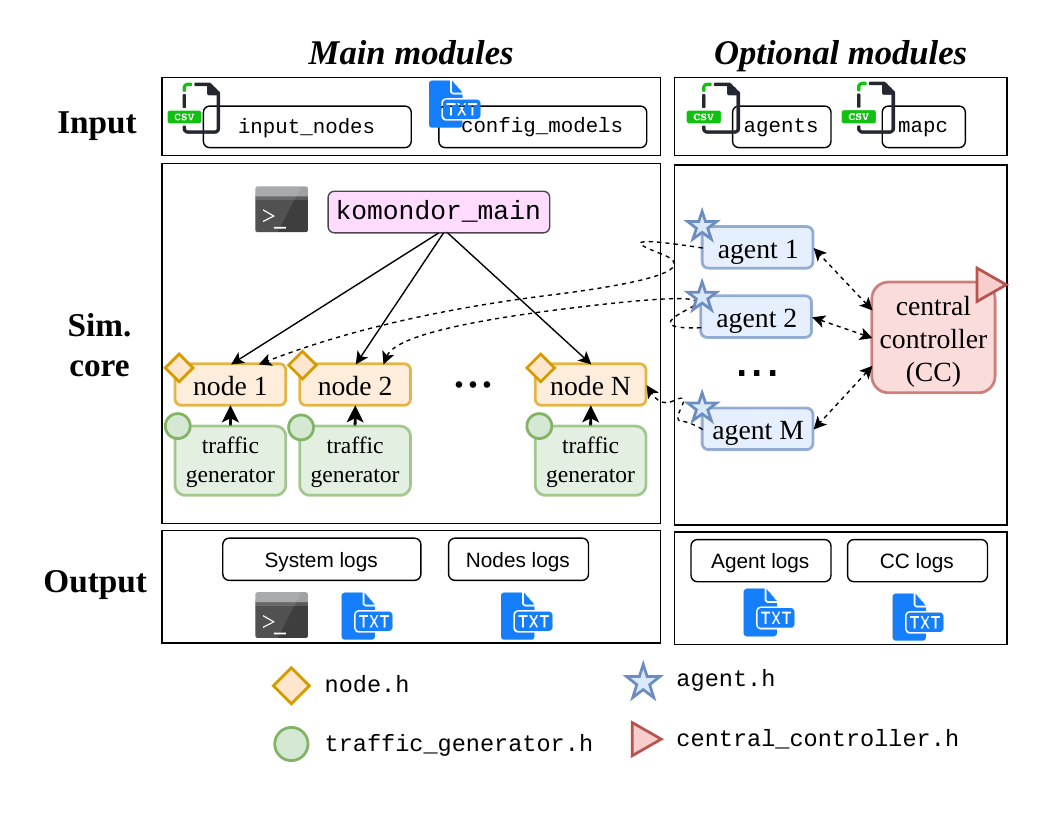}
  \caption{Kom8ndor's main components and simulation flow.}
  \label{fig:architecture}
\end{figure}

\subsection{Main simulation components}

\texttt{Kom8ndor} uses the C/C++ COST library~\cite{chen2002cost}, which allows the creation of COST components and handles discrete-event interactions among them. The main file (\texttt{komondor\_main}) inherits a \textit{CostSimEng} singleton (a special type of component defined by COST) that is in charge of instantiating all the simulation components, creating the connections between them via \textit{inport} and \textit{outport} methods, and running the simulation. \emph{TypeII} components (native to COST) are used to represent different types of entities in a Wi-Fi network, including:
\begin{itemize}
    \item \texttt{node}: Characterizes the logic of either \gls{ap} or \gls{sta} devices, including traffic management, \gls{mac}, and \gls{phy} operations.
    \item \texttt{traffic\_generator}: Generates traffic based on a selected traffic model (e.g., Poisson traffic, burst traffic, full-buffer). Every transmitting node is connected to a dedicated traffic generator.
    \item \texttt{agent}: Provides in-simulation \gls{ml} operations to optimize the performance of the network online. Agents are connected to \glspl{ap} to exchange information and commands.
    \item \texttt{central\_controller}: Orchestrates enrolled agents (e.g., centralized decision-making), thus unlocking centralized \gls{ml} and non-\gls{ml} solutions and algorithms.
\end{itemize}

The different components are flexibly instantiated based on a set of input files, comprising:
\begin{itemize}
    \item \texttt{input\_nodes}: defines per-node capabilities (e.g., position, channel, transmission power).
    \item \texttt{config\_models}: defines environment models (e.g., path loss, co-channel interference model) applied to the simulated scenario.
    \item \texttt{agents} (optional): defines \gls{ml}-related implementations, ranging from \gls{rl}-based dynamic parameter configuration to custom models interacting with \texttt{Kom8ndor} through a Python server.
    \item \texttt{mapc}  (optional): defines coordination groups, \gls{mapc} schemes, and configurations to be used by the instantiated Wi-Fi nodes.     
\end{itemize}

\texttt{Kom8ndor} outputs include simulation summaries (e.g., per-\gls{bss} performance, per-node statistics, collision counts), and detailed logs for nodes, agents, and the central controller. The overall system is illustrated in Fig.~\ref{fig:architecture}.

\subsection{Finite State Machine}
\label{subsec:node_arch}

\begin{figure}[t!]
\centering
\includegraphics[width=\linewidth]{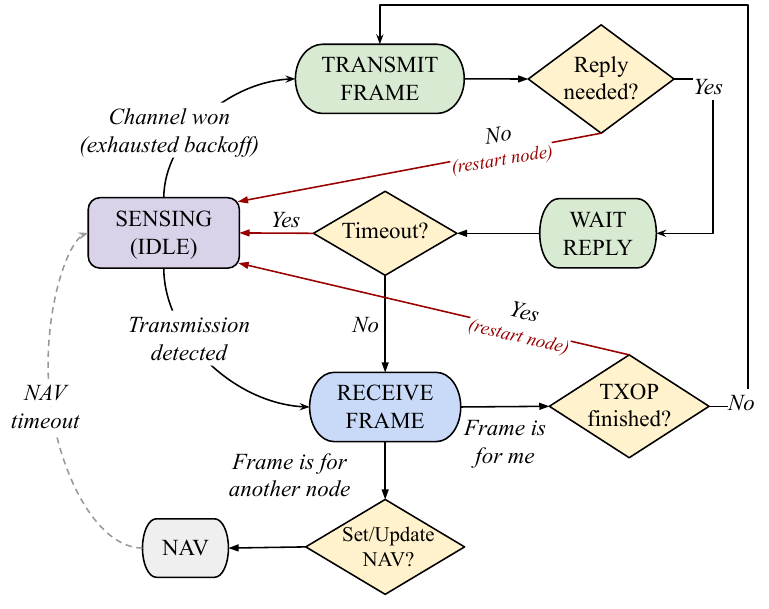}
\caption{Kom8ndor's generic FSM.}
\label{fig:fsm}
\end{figure}

\texttt{Kom8ndor} simulations are based on discrete events, following a flexible \gls{fsm}, where node states represent frame transmissions (e.g., \texttt{TRANSMIT DATA}), frame receptions (e.g., \texttt{RECEIVE DATA}), and other operations (e.g., \texttt{WAIT DATA}, \texttt{NAV}), and transitions between states follow operations such as channel contention, backoff counters, or inter-frame spacing (see Fig.~\ref{fig:fsm}). First, when a \texttt{traffic\_generator} pushes data to a \texttt{node}, the \gls{fsm} transitions through frame preparation, medium sensing, and frame transmission, scheduling events for the exact microsecond each action is expected to be performed. Accordingly, if a node has a frame to transmit, it undergoes a backoff procedure, which can run as long as the medium is sensed as idle (the node is in \texttt{SENSING} state). Upon channel access, the node transitions to the \texttt{TRANSMIT FRAME} state, which triggers \textit{transmission detected} events in other nodes that can eventually activate the \gls{nav} procedure and make them transition to \texttt{NAV} state (depending on the spatial geometry of the scenario). Depending on the need for a reply or not (e.g., \gls{ack}, \gls{cts}), the transmitting node transitions through \texttt{WAIT REPLY} and then \texttt{RECEIVE FRAME}, or goes back to \texttt{SENSING} upon finishing the transmission, respectively.

\begin{figure}[t!]
    \centering      \includegraphics[width=\columnwidth]{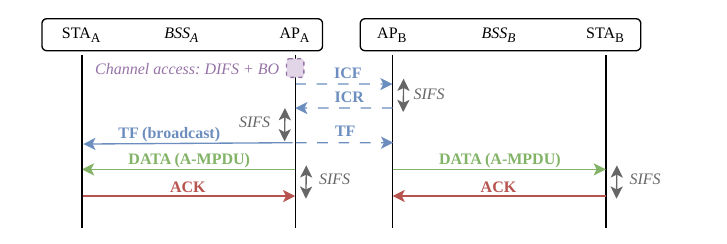}
    \caption{\gls{cosr}/\gls{cobf} frame exchange sequences.}
    \label{fig:sequences}
\end{figure}

This \gls{fsm} design provides the necessary flexibility to realize various transmission modes, ranging from simple data and \gls{ack} exchanges to more complex \gls{mapc} frame sequences. This flexibility enables seamless extensibility, allowing new transmission approaches and protocols to be easily developed. Figure~\ref{fig:sequences} shows an example of a frame sequence for \gls{cosr}/\gls{cobf}.

To support \gls{mapc} features, \texttt{Kom8ndor} includes new states and signaling that involve multiple nodes, including \gls{icf}/\gls{icr} exchange (to agree on a coordinated transmission), the transmission of a \gls{tf} (used by \gls{cosr} and \gls{cobf} to synchronize simultaneous data transmissions across multiple \glspl{ap}), or \gls{murts} (used by \gls{cotdma} to schedule time slots among coordinated \glspl{ap}). Depending on the selected \gls{mapc} scheme, the \gls{fsm} is adapted accordingly, and new transient states are introduced to manage the chronological flow of the coordination protocols. For a detailed overview of the \gls{mapc} schemes, we refer the reader to the tutorial in~\cite{wilhelmi2026mapc}.

\section{New Features: From 802.11bn to AI/ML}
\label{sec:new_features}

\subsection{Multi Access Point Coordination}

\gls{mapc} is a new feature in 802.11bn that allows a \emph{coordinating \gls{ap}} (the \gls{ap} that wins the channel access) to share its \glspl{txop} with other \emph{coordinated \glspl{ap}} and, in some cases, perform simultaneous transmissions. Current \gls{mapc} schemes comprise \gls{cotdma} (schedules multiple transmissions within the same \gls{txop}), \gls{cosr} (enables simultaneous transmissions with limited transmit power), \gls{cobf} (enables simultaneous transmissions using beamforming and nulling), \gls{cortwt} (coordinates \gls{rtwt} schedules), and \gls{cocr} (coordinates the channels used for device-to-device transmissions).

%\begin{figure}[h!]
%    \centering
%    \includegraphics[width=.8\columnwidth]{Figures/example_beamforming.pdf}
%    \caption{Scenario where two coordinated APs leverage Co-BF to perform simultaneous transmissions.}
%    \label{fig:example_beamforming}
%\end{figure}

\texttt{Kom8ndor} introduces a \gls{mapc} framework that builds upon the baseline established by 802.11bn while enabling extensions to anticipate beyond Wi-Fi~8 capabilities. For instance, while 802.11bn restricts \gls{mapc} agreements to a maximum of two \glspl{ap}, \texttt{Kom8ndor} facilitates the efficient evaluation of scalability beyond this limit. \texttt{Kom8ndor} includes the \gls{mapc} schemes that operate on a per-\gls{txop} basis, i.e., \gls{cotdma}, \gls{cosr}, and \gls{cobf}. %But once again, \texttt{Kom8ndor}'s \gls{mapc} framework allows the definition of new coordination schemes (e.g., leveraging coordination to share rewards in \gls{ml}~\cite{wilhelmi2025coordinated}).

As described in Section~\ref{sec:architecture_design}, \texttt{Kom8ndor} initializes the simulation using input files. For \gls{mapc}, agreements and parameters are provided in a new configuration file (e.g., \texttt{mapc.csv}). This file is used to pre-establish the logical coordination groups and per-scheme parameters, which depend on the scheme itself (e.g., time-split approach in \gls{cotdma}). During the simulation, \texttt{Kom8ndor} performs the new handshakes needed to coordinate shared \glspl{txop}, for which new control frames were defined (\gls{icf}, \gls{icr}, \gls{tf}, \gls{mu}-\gls{rts}). Accordingly, the \gls{fsm} is updated to support new states such as  \texttt{TRANSMIT ICF}, \texttt{WAIT ICR}, or \texttt{WAIT MU-RTS}. These states ensure that participating nodes remain synchronized and that non-participating \glspl{sta} correctly update their \gls{nav} to defer access until the end of the coordinated \gls{txop}.

%Next, we describe the specific details of each 802.11bn \gls{mapc} scheme implemented in \texttt{Kom8ndor}.

\subsubsection{Co-TDMA Operation}
The coordinating \gls{ap} splits the \gls{txop} into orthogonal time slots, assigned to different participating \glspl{ap}. After the \gls{icf}-\gls{icr} exchange, a fair \gls{txop} split is decided among coordinated \glspl{ap}, followed by the sequential data transmissions. By default, \texttt{Kom8ndor} splits the resources equally among coordinated \glspl{ap}. In particular, the per-\gls{ap} data duration $T_{n}$ for each participating \gls{ap} $n\in \mathcal{N}$ is capped equally. While not currently included, other policies, such as \gls{qos}-based scheduling or maximum-delay minimization, can be implemented.
%\begin{equation}
%T_{n} = \min \left( T^\text{req}_n, \frac{\text{TXOP}_\text{max} - T^\text{Co-TDMA}_\text{overhead}}{|\mathcal{N}|} \right),
%\end{equation}
%where $T^\text{req}_n$ is the maximum time required by \gls{ap} $n$ (depending on its buffer status and transmission parameters), $\text{TXOP}_\text{max}$ is the maximum \gls{txop} duration (5484~$\mu s$), and $T^\text{Co-TDMA}_\text{overhead}$ denotes the duration of overheads in \gls{cotdma} (including control frames and \gls{sifs} intervals). While not currently included, other policies, such as \gls{qos}-based scheduling or maximum-delay minimization, can be implemented.

\subsubsection{Co-SR Operation}
In \gls{cosr}, coordinated \glspl{ap} agree on the transmit power to be used during a simultaneous \gls{ampdu} downlink transmission. In the default implementation, the power to be used by the coordinated \gls{ap} is indicated by the user (indicated in the \texttt{mapc} input file), while the coordinating \gls{ap} employs its predefined transmit power (indicated in the \texttt{nodes} input file). However, other solutions can be implemented on top, including a smart power assignment based on the maximum allowable interference at each communication end.

\subsubsection{Co-BF Operation}
In \gls{cobf}, two \glspl{ap} transmit simultaneously thanks to coordinated inter-\gls{bss} interference suppression, where each \gls{ap} steers a beam toward its own associated \gls{sta} and places nulls in the directions of the recipient \glspl{sta} of the other \gls{bss}. \texttt{Kom8ndor}'s implementation of \gls{cobf} models each \gls{ap} as a horizontal \gls{ula} of $N$ isotropic elements\footnote{The \gls{ula} parameters ($N$, $d$, and nominal azimuth) are optionally informed in the \texttt{input\_nodes} file.} and computes beamforming weights via \gls{zf} precoding, which enforces unity gain toward the desired \gls{sta} while placing exact nulls at all $K$ peer \glspl{sta}, solving the resulting $(K{+}1)\times(K{+}1)$ Gram system via Gaussian elimination with partial pivoting. The resulting beam gain then scales the received power relative to the default omnidirectional model. Beamforming gains are applied to data frames only, while all control frames (\gls{icf}/\gls{icr}, \gls{tf}, \gls{rts}/\gls{cts}, \gls{ack}) are transmitted under the omnidirectional model.

\subsection{Flexible Spectrum Access: DSO, NPCA, and More}

\texttt{Kom8ndor} includes two new spectrum access features: \gls{dso} and \gls{npca}. In addition, Komondor's \gls{dcb} baseline is extended~\cite{barrachina2019dynamic}.

\subsubsection{Dynamic Subband Operation (DSO)}
\gls{dso} tries to fill the increasing capability gap between \gls{ap} and \gls{sta} devices, where \glspl{ap} might use wide channels (e.g., 160~MHz or 320~MHz), while \glspl{sta} might have more moderate capabilities (e.g., 20~MHz or 40~MHz). With \gls{dso}, narrow-band \glspl{sta} can be allocated to secondary subchannels during the same \gls{txop} in a wide-band transmission, hence increasing efficiency. For example, without \gls{dso}, a 160~MHz \gls{ap} serving four 40~MHz \glspl{sta} would use only 40~MHz per \gls{txop}, wasting 120~MHz each time. With \gls{dso}, in contrast, the four \glspl{sta} could be scheduled in a single \gls{txop}. \texttt{Kom8ndor} introduces \gls{dso} by simulating the mandatory \gls{icf}/\gls{icr} exchange before the data transmission begins. After that, all the scheduled \glspl{sta} transmit on their assigned portion of the bandwidth. Currently, the set of scheduled \glspl{sta} in a \gls{dso} transmission is decided using round-robin, allocating in order the selected \glspl{sta} based on their declared bandwidth (specified as \texttt{min\_ch}/\texttt{max\_ch} in the \texttt{input\_nodes} file).

\subsubsection{Non-Primary Channel Access (NPCA)}
\gls{npca} aims to overcome the contention aspects in an \gls{obss} by allowing an \gls{ap} to temporarily translate its operation to a different channel. When an \gls{npca} node detects an inter-\gls{bss} transmission, it switches to a pre-configured \gls{npca} primary channel that is within the total \gls{bss} bandwidth. There, the node can run a new backoff (as if it was the primary) and eventually initiate a transmission. At the end of the \gls{txop}, the node must return to the primary channel. \texttt{Kom8ndor} implements \gls{npca} by capturing its sequence of events. First, the detection of an \gls{obss} transmission triggers the \gls{npca} channel switching (a fixed configurable radio switching delay can be added), where a new backoff is run. Then, a transmission can occur on the \gls{npca} primary channel, which must be preceded by an \gls{icf}/\gls{icr} exchange. Regardless of whether a transmission is performed or not on the \gls{npca} primary channel, an \gls{npca} trigger determines the time at which the node returns to its primary channel. The \gls{npca} primary channel can be flexibly defined in \texttt{Kom8ndor}, but it defaults to the upper portion of the \gls{bss} bandwidth.

\subsubsection{Preamble Puncturing}
Based on the \gls{dcb} implementation in Komondor, \texttt{Kom8ndor} adds \emph{preamble puncturing}, which was introduced in IEEE~802.11ax. In preamble puncturing, the \gls{txop} remains anchored to the primary subband, but spectrum is aggregated more flexibly. In particular, any individual busy secondary 20~MHz channels are excluded from the set of channels bonded for a data transmission (achieved by marking the busy basic channels in a puncturing bitmap), making the most of the available bandwidth.

\subsection{Advanced Channel Access Methods}

Channel access has traditionally been an object of discussion in Wi-Fi. The standard \gls{dcf} relies on \gls{csmaca} with \gls{beb}, whereby nodes initiate transmissions upon successfully decrementing a random backoff while the channel is idle; despite its simplicity and long adoption, \gls{dcf} inefficiencies are well known, especially as the number of coexisting nodes increases. Beyond baseline \gls{dcf}, \texttt{Kom8ndor} supports \gls{edca}~\cite{bellalta2006modeling}, today's baseline mechanism, which extends \gls{dcf} with \gls{qos} differentiation across four \glspl{ac}---\gls{vo}, \gls{vi}, \gls{be}, \gls{bk}---and is fully configurable per node and \gls{ac} to set the \gls{aifs}, \gls{cw} bounds, and \gls{txop} limit. It also supports several determinism-oriented alternatives: \emph{deterministic backoff}~\cite{tinnirello2025deterministic}, which fixes $BO = b + \text{IPT}$ (a base value plus the number of backoff interruptions suffered) and reverts to random backoff only after collisions; \gls{iyt}~\cite{wilhelmi2025s}, a distributed round-robin scheme where a logical token is passed to the next node in a shared transmission-order list after each success; and simpler heuristic baselines such as \gls{eca}~\cite{sanabria2016high} (repeats the same backoff after a success), Repeat Backoff, and Synchronized Backoff (fixed, non-adaptive backoff), useful as analytical references for studying the limits of channel access.

\subsection{Machine Learning Tools}
\label{subsec:ml_agents}

Since its first release in 2018, Komondor has included \gls{ai} modules for studying next-generation \glspl{wlan}. Specifically, \texttt{agent} and \texttt{central\_controller} components were introduced to realize \gls{rl} algorithms to dynamically optimize operations such as channel allocation, \gls{sr}, or \gls{dcb}. Now, \texttt{Kom8ndor} has improved this design and has added, on top, a seamless integration with Python, which is key to leveraging third-party \gls{ml} packages such as PyTorch~\cite{paszke2019pytorch}. 

\subsubsection{Built-in Learning Algorithms}

\texttt{Kom8ndor} includes built-in learning functions that allow the dynamic configuration of Wi-Fi \gls{phy} and \gls{mac} parameters during the simulation time. In particular, a single dispatch point handles all the agent decisions. Currently, a \gls{mab} module is implemented to enable online decision-making, in which the agent maintains empirical reward estimates for each arm (e.g., a channel configuration) and selects among them via a configurable exploration strategy (e.g., $\varepsilon$-greedy). Other implementations are easily pluggable, since algorithmic differences between \gls{rl} paradigms (e.g., model-based vs model-free) are hidden from the rest of the simulator. With this architecture, different learning paradigms can be flexibly instantiated: \emph{i) Decentralized} (agents perform decision-making based on local information only~\cite{wilhelmi2026decentralized}), \emph{ii) Coordinated} (multiple agents exchange information to perform decision-making cooperatively~\cite{wilhelmi2025coordinated}), and \emph{iii) Centralized} (a central controller gathers global state information from agents and performs centralized decision-making).
%\begin{itemize}
    %\item \textbf{Decentralized learning:} Agents are directly attached to a \gls{bss}, from which they continuously gather telemetry data (e.g., achieved throughput, collision rates, average delay, and observed interference). Individual decision-making is made based on local information only~\cite{wilhelmi2019potential, wilhelmi2026decentralized}. 
    %
    %\item \textbf{Coordinated learning:} Multiple agents exchange information (e.g., rewards) to perform decision-making cooperatively~\cite{wilhelmi2025coordinated}.
   %\item \textbf{Centralized orchestration:} Agents connect to a central controller, which gathers global state information and performs centralized decision-making, allowing the simulator to characterize cloud-based \gls{ai} orchestrators.
%\end{itemize}

\subsubsection{External Python Integration}
To run complex \gls{ml} paradigms such as \glspl{dnn} and leverage existing libraries such as PyTorch, TensorFlow, or scikit-learn, \texttt{Kom8ndor} introduces a new agent mode that offloads decisions to an external Python process. With this, agents can send a feature vector at each step or iteration to inform about the played action index, the current configuration, or the observed reward (performance). While the current implementation remains simple, it can be easily extended to include richer state representations (e.g., channel measurements, queue status information). The communication between the simulator and the Python process uses a Portable Operating System Interface (POSIX) Unix-domain socket. In particular, in each iteration, \texttt{Kom8ndor} sends a 32-bit integer header specifying the number of features, followed by a vector of 32-bit floats with the values of those features. The Python server returns a vector of 32-bit floats encoding the \gls{ml} model output (e.g., next action).

\section{Tutorial \& Performance Showcase}
\label{sec:simulations}

\begin{table}[t!]
\centering
\caption{\texttt{Kom8ndor} simulation flags.}
\label{tab:flags}
\resizebox{\columnwidth}{!}{%
\begin{tabular}{@{}llll@{}}
\toprule
 & Flag & Arg & Description \\ \midrule
\multirow{5}{*}{\begin{tabular}[c]{@{}l@{}}Sim. \\ params.\end{tabular}} & \texttt{--nodes} (\texttt{-n}) & file & Path to nodes file \\
 & \texttt{--time} (\texttt{-t}) & float & Simulation duration in seconds \\
 & \texttt{--seed} (\texttt{-s}) & int & Random seed \\
 & \texttt{--code} (\texttt{-c}) & str & Simulation code/ID \\
 & \texttt{--out} (\texttt{-o}) & file & Path to output file \\ \midrule
\multirow{2}{*}{Advanced} & \texttt{--mapc} (\texttt{-m}) & file & Path to MAPC file \\
 & \texttt{--agents} (\texttt{-a}) & file & Path to agents file \\ \midrule
\multirow{4}{*}{Logging} & \texttt{--logs-sys} (\texttt{-L}) & 0/1 & Print system events to terminal \\
 & \texttt{--logs-node} (\texttt{-l}) & 0/1 & Print per-node events to terminal \\
 & \texttt{--save-node} (\texttt{-S}) & 0/1 & Write per-node logs to file \\
 & \texttt{--save-agent} (\texttt{-A}) & 0/1 & Write per-agent logs to file \\ \bottomrule
\end{tabular}%
}
\end{table}

\texttt{Kom8ndor} is a command-line simulator that runs from the terminal or through multi-scenario shell scripts. Running \texttt{Kom8ndor} requires a C++ compiler (g++ with C++98 support) and a Linux or macOS environment (or WSL on Windows) to run the COST code-generation step. A \textit{Makefile} is included in \texttt{Kom8ndor} to automatically build the source code (including the necessary COST libraries) and generate the binary \texttt{komondor\_main} that runs the simulation. The \textit{Makefile} also tracks all \textit{*.cc} and \textit{*.h} files as dependencies, so any change to any header (including the ones in Code/methods/) triggers a full rebuild. To build the code, one must run
\begin{bashbox}{Build the code}
cd Code/main && make
\end{bashbox}

Simulations are launched by invoking \texttt{komondor\_main} with the necessary flags to specify the scenario and running features (\texttt{--nodes}, \texttt{--mapc}, \texttt{--agents}) together with simulation parameters (\texttt{--time}, \texttt{--seed}, \texttt{--code}). Log verbosity is controlled via \texttt{--logs-sys} and \texttt{--logs-node}, and output results are written to a generic .txt file (\texttt{--out}) and individual per-node and per-agent files (\texttt{--save-node}, \texttt{--save-agent}).\footnote{Saving logs entails high simulation times, particularly when simulating large deployments for a long time.} The complete list of \texttt{Kom8ndor} flags is provided in Table~\ref{tab:flags}.

\subsection{Running MAPC Schemes}
\label{sec:tutorial_mapc}

We consider an example of a simulation involving \gls{mapc} schemes, where two overlapping \glspl{bss} ($A$ and $B$) share the same $40$~MHz channel. We run four different cases: \emph{i)} \gls{dcf}, \emph{ii)} \gls{cotdma} (an equal time split between \glspl{ap} is considered), \emph{iii)} \gls{cosr} (the coordinating \gls{ap} uses full power while the coordinated \gls{ap} power is limited to $15$~dBm), and \emph{iv)} \gls{cobf} (an 8-element \gls{ula} is considered for beamforming).\footnote{The different input files used across this section can be found in folder \url{https://github.com/wn-upf/Komondor/tree/master/Code/input/examples}.} The following commands are used to run these scenarios: 

\begin{bashbox}{Running MAPC examples}
# DCF only
./komondor_main -n ../input/examples/mapc_
example/input_nodes_mapc_example.csv 
-t 100 -s 1

# Scenario with MAPC (Co-TDMA example)
./komondor_main -n ../input/examples/mapc_
example/input_nodes_mapc_example.csv -m
../input/examples/mapc_example/mapc_cotdma_
example.csv -t 100 -s 1
\end{bashbox}

\begin{figure}[b!]
    \centering    
%    \subfloat[]{\includegraphics[width=.8\columnwidth]{Figures/scenario_mapc.pdf}\label{fig:scenario_mapc}}
%    \par\vspace{0.2cm}
    \includegraphics[width=.85\columnwidth]{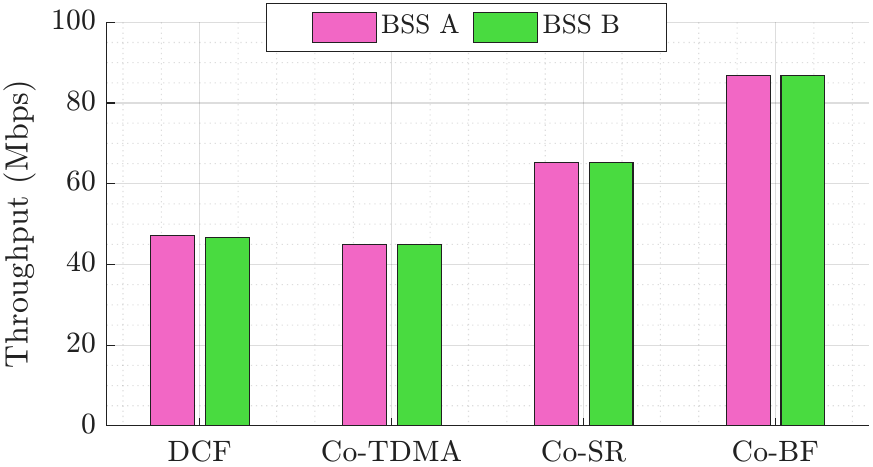}
    \caption{Throughput achieved per STA in the MAPC example.}
\label{fig:mapc_example}
\end{figure}

The resulting throughput per \gls{bss} and scheme, which were extracted from \texttt{Kom8ndor} logs, are plotted in Fig.~\ref{fig:mapc_example}. In this simple scenario, \gls{cotdma} leads to slightly lower throughput than \gls{dcf} because of the overheads associated with the coordination. \gls{cosr} increases the throughput further thanks to the geometric position of nodes in this deployment, which favors space reutilization. Finally, \gls{cobf} shows the greatest improvement due to the \gls{sinr} gains that result from coordinated nulling.

%\begin{figure}[t!]
%    \centering    
%    \subfloat[]{\includegraphics[width=.8\columnwidth]{Figures/scenario_mapc.pdf}\label{fig:scenario_mapc}}
%    \par\vspace{0.2cm}
%    \subfloat[]{\includegraphics[width=.85\columnwidth]{Figures/throughput_mapc_example.pdf}\label{fig:throughput_mapc_example}}
%    \caption{MAPC example. (a) Scenario, (b) Throughput achieved per STA and scheme.}
%    \label{fig:mapc_example}
%\end{figure}

\subsection{Running NPCA and DSO}
\label{sec:tutorial_dsonpca}

We consider a single deployment with two \glspl{bss} to show \gls{dso} and \gls{npca} operation. In particular, \gls{bss} $A$ uses a $40$~MHz channel and has two associated $20$-MHz \glspl{sta} with \gls{dso} enabled. \gls{bss} $B$, instead, uses $80$~MHz, with the first $40$~MHz portion overlapping with \gls{bss} $A$'s bandwidth. To run these scenarios, the following commands are used:

\begin{bashbox}{Running DSO and NPCA}
# DCF
./komondor_main -n ../input/examples/
dso_npca_example/input_nodes_dcf.csv 
-t 100 -s 1

# DSO and NPCA
./komondor_main -n ../input/examples/
dso_npca_example/input_nodes_dso_npca.csv 
-t 100 -s 1
\end{bashbox}

\begin{figure}[b!]
    \centering    
    \includegraphics[width=.85\columnwidth]{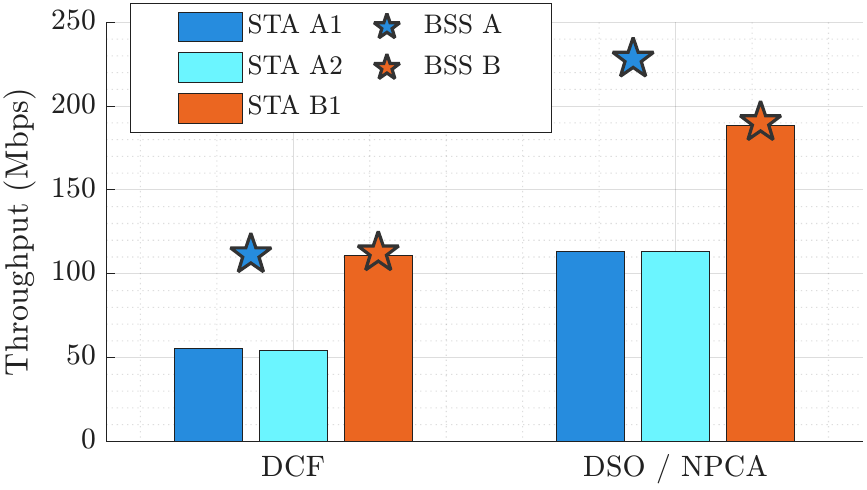}
    \caption{Throughput achieved per STA/BSS in the DSO \& NPCA example.}
    \label{fig:dso_npca_example}
\end{figure}

The resulting throughput is shown in Fig.~\ref{fig:dso_npca_example}: The two \glspl{bss} get the same aggregate throughput (equally split between the two \glspl{sta} in \gls{bss} $A$). With \gls{dso} (\gls{bss} $A$) and \gls{npca} (\gls{bss} $B$), significant gains are observed, which result from the higher flexibility of the two \glspl{bss} in accessing the medium, where \gls{sta} $A1$ and $A2$ can be served simultaneously by \gls{ap} $A$ on channels \#$0$ and \#$1$, respectively, and \gls{bss} $B$ can access an unused portion of the spectrum (channels \#$2$ and \#$3$).

%\begin{figure}[t!]
%    \centering    
%    \subfloat[]{\includegraphics[width=.8\columnwidth]{Figures/scenario_dso_npca.pdf}\label{fig:scenario_dso_npca}}
%    \par\vspace{0.2cm}
%    \subfloat[]{\includegraphics[width=.85\columnwidth]{Figures/throughput_dso_npca_example.pdf}\label{fig:throughput_dso_npca_example}}
%    \caption{DSO \& NPCA example. (a) Scenario, (b) Throughput achieved per STA, BSS, and scheme.}
%    \label{fig:dso_npca_example}
%\end{figure}

\subsection{Running ML Agents with Python}
\label{sec:tutorial_ml}

When running the newly added external \gls{ml} models (indicated as \texttt{learning mechanism=7} in the \texttt{agents} input), the agent delegates arm selection (or other tasks) to an external Python server. The following example (\texttt{ml\_server\_random.py}) simply selects random actions:

\begin{bashbox}{Running ML with Python}
# Terminal 1: start the server first
python3 Code/learning_modules/python_servers/
ml_server_random.py /tmp/komondor_ml.sock

# Terminal 2: run the simulation (Code/main)
./komondor_main -n ../input/examples/
external_model_example/input_nodes.csv 
-a ../input/examples/external_model_example/
agents_external.csv -t 100 -s 1
\end{bashbox}

%On the server side, it is displayed the information of each interaction (current arm, reward, and next selected arm):

%\begingroup
%\scriptsize
%\begin{verbatim}
%[ml_server_random] Listening on /tmp/komondor_ml.sock ...
%[ml_server_random] Client #1 connected.
%[ml_server_random] Client #2 connected.
%  recv: arm_ix=0  reward=0.4261  num_arms=2  next_arm=1
%  recv: arm_ix=0  reward=0.5351  num_arms=2  next_arm=0
%  recv: arm_ix=1  reward=0.0000  num_arms=2  next_arm=0
%  ...
%\end{verbatim}
%\endgroup

%On the \texttt{Kom8ndor} side, the simulation takes place, showing at the end the summary for each agent:

%\begingroup
%\scriptsize
%\begin{verbatim}
%------- Agent A0 ------
%    · External model: socket=/tmp/komondor_ml.sock 
%    · Cumulative reward per arm: 49.325949  0.000000  
%    · Times each arm has been selected: 105  94  
%
%------- Agent A1 ------
%    · External model: socket=/tmp/komondor_ml.sock
%    · Cumulative reward per arm: 47.210786  0.010393  
%    · Times each arm has been selected: 99  100  
%\end{verbatim}
%\endgroup

%While not shown here, two more examples are provided in \texttt{Code/learning\_modules/}: \texttt{ml\_server\_passthrough.py} (echoes the current arm index back, to verify the socket connection) and \texttt{ml\_server\_pytorch.py} (loads a pre-trained model). In all the cases, the socket path is configured through the \texttt{agents} files and defaults to \texttt{/tmp/komondor\_ml.sock}. 
Beyond \gls{mab} algorithms, the external server enables countless Python implementations (\texttt{ml\_server\_pytorch.py} uses a pre-trained \gls{dnn} for inference). The same interface naturally supports algorithms like \gls{dqn}, turning \texttt{Kom8ndor} into an \gls{rl} gym like~\cite{gawlowicz2019ns}.

\section{Conclusions and Future Work}
\label{sec:conclusions}

\texttt{Kom8ndor} is a live project. It is the basis for simulation-driven research on Wi-Fi 8 and beyond in the Wireless Networking research group at Universitat Pompeu Fabra (UPF). In this paper, we present the first release of Wi-Fi 8, outlining its main features so far: \gls{mapc}, \gls{npca}, \gls{dso}, and the integration of \gls{ml} tools. However, this does not stop here, and more work is in the pipeline. First, relevant features from previous generations of Wi-Fi, e.g., \gls{mlo}, \gls{ofdma}, or \gls{rtwt}, will be included to study their interplay with Wi-Fi 8 functionalities. In addition, beyond Wi-Fi 8 \gls{mapc} solutions will be progressively included and evaluated, such as a dynamic transmit power computation for \gls{cosr}, \gls{qos}-based \gls{cotdma} scheduling, or smart \gls{mapc} scheme selection.

%%%%%%%%%%%%%%%%%%%%%%%%%
%%%  BIBLIOGRAPHY    
%%%%%%%%%%%%%%%%%%%%%%%%%
%\bibliographystyle{unsrt}
\bibliographystyle{IEEEtran}
\bibliography{references}

@IEEEtranBSTCTL{IEEEexample:BSTcontrol,
CTLuse_forced_etal       = "yes",
CTLmax_names_forced_etal = "3",
CTLnames_show_etal       = "1",
CTLdash_repeated_names = "no"
}

@inproceedings{barrachina2019komondor,
  title={{Komondor: A wireless network simulator for next-generation high-density WLANs}},
  author={Barrachina-Munoz, Sergio and Wilhelmi, Francesc and Selinis, Ioannis and Bellalta, Boris},
  booktitle={Wireless Days},
  pages={1--8},
  year={2019},
  organization={IEEE}
}

@article{wilhelmi2021spatial,
  title={{Spatial Reuse in IEEE 802.11 ax WLANs}},
  author={Wilhelmi Roca, Francesc and Barrachina-Mu{\~n}oz, Sergio and Cano, Cristina and Selinis, Ioannis and Bellalta, Boris},
  journal={Comp. Comm. 2021, 170: 65-83},
  year={2021},
  publisher={Elsevier}
}

@inproceedings{wilhelmi2025s,
  title={{It's Your Turn: A Novel Channel Contention Mechanism for Improving Wi-Fi's Reliability}},
  author={Wilhelmi, Francesc and Galati-Giordano, Lorenzo and Fontanesi, Gianluca},
  booktitle={WCNC},
  pages={1--6},
  year={2025},
  organization={IEEE}
}

@inproceedings{bellalta2006modeling,
  title={{Modeling the IEEE 802.11 e EDCA for MAC parameter optimization}},
  author={Bellalta, Boris and Cano, Cristina and Oliver, Miquel and Meo, Michela},
  booktitle={Het-Nets},
  year={2006}
}

@article{nunez2025enabling,
  title={{Enabling reliable latency in Wi-Fi 8 through multi-AP joint scheduling}},
  author={Nunez, David and Imputato, Pasquale and Avallone, Stefano and Smith, Malcolm and Bellalta, Boris},
  journal={OJCOMS},
  year={2025},
  publisher={IEEE}
}

@inproceedings{carrascosa2024performance,
  title={{Performance evaluation of MLO for XR streaming: Can Wi-Fi 7 meet the expectations?}},
  author={Carrascosa-Zamacois, Marc and Galati-Giordano, Lorenzo and Wilhelmi, Francesc and Fontanesi, Gianluca and Jonsson, Anders and Geraci, Giovanni and Bellalta, Boris},
  booktitle={CAMAD},
  pages={1--6},
  year={2024},
  organization={IEEE}
}

@inproceedings{alsakati2023performance,
  title={{Performance of 802.11 be Wi-Fi 7 with multi-link operation on AR applications}},
  author={Alsakati, Molham and Pettersson, Charlie and Max, Sebastian and Moothedath, Vishnu Narayanan and Gross, James},
  booktitle={WCNC},
  pages={1--6},
  year={2023},
  organization={IEEE}
}

@article{liu2024wi,
  title={{Wi-Fi 8: Embracing the millimeter-wave era}},
  author={Liu, Xiaoqian and Chen, Tingwei and Dong, Yuhan and Mao, Zhi and Gan, Ming and Yang, Xun and Lu, Jianmin},
  journal={IEEE Commun. Mag.},
  volume={63},
  number={3},
  pages={69--75},
  year={2024},
  publisher={IEEE}
}

@article{jeknic2023development,
  title={{Development steps that brought to Wi-Fi 7}},
  author={Jekni{\'c}, Ana and Ko{\v{c}}an, Enis},
  journal={ETF Journal of Electrical Engineering},
  volume={29},
  number={1},
  pages={65--79},
  year={2023}
}

@article{chen2022overview,
  title={{Overview and performance evaluation of Wi-Fi 7}},
  author={Chen, Cheng and Chen, Xiaogang and Das, Dibakar and Akhmetov, Dmitry and Cordeiro, Carlos},
  journal={IEEE Commun. Standards Mag.},
  volume={6},
  number={2},
  pages={12--18},
  year={2022},
  publisher={IEEE}
}

@inproceedings{chen2002cost,
  title={{Cost: A component-oriented discrete event simulator}},
  author={Chen, Gilbert and Szymanski, Boleslaw K},
  booktitle={Proc. of the Winter Simulation Conf.},
  volume={1},
  pages={776--782},
  year={2002},
  organization={IEEE}
}

@article{wilhelmi2026mapc,
  title={{A Tutorial on IEEE 802.11bn Multi-AP Coordination for Wi-Fi 8: From Standardization to Performance Evaluation}},
  author={Wilhelmi, Francesc and Bellalta, Boris and Geraci, Giovanni and Galati-Giordano, Lorenzo and Meneghello, Francesca and Kijanka, Aleksandra and Val, Iñaki and López-Pérez, David},
  journal={arXiv preprint arXiv:2606.13759},
  year={2026},
}

@article{geraci2026wi,
  title={{Wi-Fi: 25 Years and Counting}},
  author={Geraci, Giovanni and Meneghello, Francesca and Wilhelmi, Francesc and L{\'o}pez-P{\'e}rez, David and Val, Inaki and Giordano, Lorenzo Galati and Cordeiro, Carlos and Ghosh, Monisha and Knightly, Edward and Bellalta, Boris},
  journal={Proc. IEEE},
  year={2026},
  publisher={IEEE}
}

@inproceedings{wilhelmi2025coordinated,
  title={{Coordinated multi-armed bandits for improved spatial reuse in Wi-Fi}},
  author={Wilhelmi, Francesc and Bellalta, Boris and Szott, Szymon and Kosek-Szott, Katarzyna and Barrachina-Mu{\~n}oz, Sergio},
  booktitle={ICMLCN},
  pages={1--6},
  year={2025},
  organization={IEEE}
}

@article{tinnirello2025deterministic,
  title={{A Deterministic Backoff Approach for Wi-Fi and NR-U Coexistence in Shared Bands}},
  author={Tinnirello, Ilenia and Wentink, Menzo and Valvo, Alice Lo and Szott, Szymon and Kosek-Szott, Katarzyna},
  journal={IEEE Trans. Mobile Comp.},
  year={2025},
  publisher={IEEE}
}

@article{sanabria2016high,
  title={{A high efficiency MAC protocol for WLANs: Providing fairness in dense scenarios}},
  author={Sanabria-Russo, Luis and Barcelo, Jaume and Bellalta, Boris and Gringoli, Francesco},
  journal={IEEE/ACM Trans. on Networking},
  volume={25},
  number={1},
  pages={492--505},
  year={2016},
  publisher={IEEE}
}

@article{barrachina2019dynamic,
  title={{Dynamic channel bonding in spatially distributed high-density WLANs}},
  author={Barrachina-Munoz, Sergio and Wilhelmi, Francesc and Bellalta, Boris},
  journal={IEEE Trans. Mobile Comp.},
  volume={19},
  number={4},
  pages={821--835},
  year={2019},
  publisher={IEEE}
}

@article{paszke2019pytorch,
  title={{Pytorch: An imperative style, high-performance deep learning library}},
  author={Paszke, Adam and Gross, Sam and Massa, Francisco and Lerer, Adam and Bradbury, James and Chanan, Gregory and Killeen, Trevor and Lin, Zeming and Gimelshein, Natalia and Antiga, Luca and others},
  journal={NeurIPS},
  volume={32},
  year={2019}
}

@inproceedings{wilhelmi2026decentralized,
  title={{Decentralized Spatial Reuse Optimization in Wi-Fi: An Internal Regret Minimization Approach}},
  author={Wilhelmi, Francesc and Bellalta, Boris and Casasnovas, Miguel and Kijanka, Aleksandra and Calvo-Fullana, Miguel},
  booktitle={EuCNC/6G Summit},
  pages={254--259},
  year={2026},
  organization={IEEE}
}

@incollection{riley2010ns,
  title={{The ns-3 network simulator}},
  author={Riley, George F and Henderson, Thomas R},
  booktitle={Modeling and tools for network simulation},
  pages={15--34},
  year={2010},
  publisher={Springer}
}

@inproceedings{gawlowicz2019ns,
  title={{Ns-3 meets openai gym: The playground for machine learning in networking research}},
  author={Gaw{\l}owicz, Piotr and Zubow, Anatolij},
  booktitle={MSWiM},
  pages={113--120},
  year={2019}
}

@inproceedings{varga2008overview,
  title={{An overview of the OMNeT++ simulation environment}},
  author={Varga, Andr{\'a}s and Hornig, Rudolf},
  booktitle={SIMUtools},
  pages={1--10},
  year={2008}
}

@misc{matlab_wlan_toolbox,
  author       = {MathWorks},
  title        = {{WLAN Toolbox}},
  note         = {{R2026a}},
  url          = {https://www.mathworks.com/products/wlan.html}
}

\end{document}